\documentstyle[12pt]{article}
\def\beq{\begin{equation}}
\def\brr{\begin{array}}
\def\err{\end{array}}
\def\ben{\begin{enumerate}}
\def\een{\end{enumerate}}
\def\eeq{\end{equation}}
\def\bea{\begin{eqnarray}}
\def\eea{\end{eqnarray}}
\def\bs{\bigskip}
\def\tr{\mbox{Tr}\, }
\def\ni{\noindent}
\def\wt{\widetilde}
\def\wh{\widehat}

\def\nn{\nonumber}
\def\ms{\medskip}

\def\txs{\textstyle}
\def\dsp{\displaystyle}

\begin{document}

\HFILL UB-ECM-PF 92/    \MBOX{}

\vspace*{1cm}

\begin{center}

{\LARGE \bf
One-loop renormalization in two-dimensional
matter-dilaton quantum gravity and charged black holes}

\vspace{8mm}

{\sc E. Elizalde\footnote{\it e-mail: eli @ ebubecm1.bitnet} and
S.D. Odintsov}\footnote{On leave from
Department of
Mathematics and Physics,
 Pedagogical Institute, 634041 Tomsk, Russia.}
\ms

Department E.C.M., Faculty of Physics, \\
University of Barcelona, \\
Diagonal 647, 08028 Barcelona, Spain

\vspace{1cm}

{\sl June 1992}

\vspace{1cm}

{\bf Abstract}

\end{center}

The quantum properties of two-dimensional matter-dilaton gravity
 ---which includes a large family of actions for two-dimensional
gravity (in particular, string-inspired models)--- are
investigated. The one-loop divergences in  linear covariant
gauges are calculated and the structure of the one-loop
renormalization is studied. The explicit forms of the dilaton
potential, dilaton-Maxwell, and dilaton-scalar couplings for which
the theory is one-loop multiplicatively renormalizable are found.
 A comparison with the one-loop renormalization structure of
four-dimensional gravity-matter theory is given. Charged
multiple-horizon black holes which appear in the model are also
considered.
\vspace{1cm}


\newpage

\section{Introduction.}

Witten's identification of a black hole solution in string theory
[1] has originated a considerable activity in the discussion of
black holes in two-dimensional gravity [2-7]. For instance, the
multiplicatively renormalizable and classically soluble model of
two-dimensional gravity (i.e., scalar-tensor or dilaton gravity)
with matter has been investigated in ref. [3] as a toy model for
the evaporation and formation of black holes [8,9]. The properties
of black holes in this model have been further investigated in
refs. [4-7]. In all, dilaton gravity  provides us with the
possibility to
have a very useful toy model, which is still complicated enough,
and which can tell us a lot about the general properties of quantum
gravity.
The study of the quantum properties of two-dimensional dilaton
gravity [10-16] shows that for some choices of the dilaton
potential the theory can be perturbatively renormalizable [12-16],
or even finite [15]. This is in contrast with four-dimensional
Einstein gravity, which is one-loop finite on shell (or off shell,
in some gauges) but which, unfortunately, is  not perturbatively
renormalizable. It is therefore quite natural to study now the
quantum aspects of dilaton gravity with matter in the perturbative
approach. Matter can be described by scalar fields, as in refs.
[3-7], or by scalar and vector fields, as in [17,18].

Our main purpose in the present paper will be to consider the
quantum corrections (one-loop renormalization structure) of
two-dimensional matter-dilaton quantum gravity with
the following action
\beq
S=\int d^2x \, \sqrt{g} \, e^{-2\phi} \left[ R+ \gamma g^{\mu\nu}
\partial_{\mu} \phi \partial_{\nu}  \phi -\frac{1}{4} f
(\phi) F_{\mu\nu}^2 + b(\phi ) g^{\mu\nu}
\partial_{\mu} \psi \partial_{\nu}  \psi+ V( \phi, \psi ) \right],
\eeq
where  $F_{\mu\nu} =\partial_{\mu} A_{\nu}-\partial_{\nu}
A_{\mu}$ is the electromagnetic field-strength, $\phi$ the
dilaton, $f(\phi )$ and  $b(\phi )$ arbitrary functions of the
dilaton field $\phi$, and $V(\phi, \psi )$ a general potential.
The first two terms in (1) represent the action of dilaton gravity,
the third one is the Maxwell term interacting with gravity via the
dilaton-Maxwell coupling $f(\phi )$, and the fourth term gives the
action of the scalar, 'spectator' field $\psi$ interacting with
gravity via the scalar-dilaton coupling $b(\phi )$. Notice that the
dilaton potential $V$ has dimensions of $[V]= M^{-2}$. The scalars
$\psi$ appear in the superstring context from the Ramond-Ramond
field.

The model with the action (1) is connected with four or
higher-dimensional Einstein-Maxwell theories (and their
generalizations), which admit charged black hole solutions [19,20].
It can be obtained from these theories via some compactification
[18].
The theory with the action (1) can also be considered as a toy
model for four-dimensional gravity-matter theory, i.e., as a
two-dimensional analog of gravity plus grand unification theory.
 Finally, as it has been
shown in refs. [17,18], for
$b(\phi) =$ const.,
$f(\phi) =$ const.
 and a dilaton
potential of the type produced by string loops, the theory admits
charged black holes with multiple horizons.

Note also that particular cases of action (1) describe a number of
well-known models, like the Jackiw-Teitelboim model [10]
($A_{\mu}= \psi =\gamma =0$, $V(\phi )=\Lambda$),
the bosonic string effective action
($A_{\mu}= \psi =0$, $\gamma =4$, $V(\phi )=\Lambda$),
and the heterotic string effective action
($f(\phi )=1$, $b(\phi )=$ const., $V(\phi )=\Lambda$).

The paper is organized as follows. In section 2 we calculate the
one-loop divergences in the Maxwell-dilaton gravity sector of
the theory (1) (with a slightly modified action). In section 3 we
complete the calculation of the divergences of the theory, by
taking into account the scalar-dilaton  gravity sector
contributions. The structure of one-loop renormalization is
discussed in section 4. In particular, we show that the theory
can be one-loop multiplicatively renormalizable for some specific
choices
of the dilaton potential and of the dilaton-Maxwell and
dilaton-scalar couplings. A comparison with the one-loop
renormalization of the four-dimensional matter-gravity  system is
also performed.
In section 5 the classical equations of motion and charged black holes
of the theory are discussed. Finally, section 6 is devoted to
conclusions.
 \bs

\section{One-loop divergences: The Maxwell sector contributions.}

Let us start from the action (1). We shall write this action in a
slightly different form which is more convenient for the
investigation of its renormalization structure:
\beq
S = \int d^2x \, \sqrt{g} \, e^{-2\phi} \left[ \frac{1}{2}
g^{\mu\nu} \partial_{\mu} \phi \partial_{\nu} \phi + \gamma R -
\frac{1}{4} f(\phi ) F_{\mu\nu}^2
+ b(\phi ) g^{\mu\nu} \partial_{\mu} \psi \partial_{\nu}
\psi + V(\phi, \psi) \right].
\eeq
This action will now be our starting point. With the change of
field variable: \, $\wt{\phi} = e^{-\phi}$,  we can write it in
the following way
\bea
S &=& \int d^2x \, \sqrt{g} \,  \left[ \frac{1}{2}
g^{\mu\nu} \partial_{\mu} \wt{\phi} \partial_{\nu} \wt{\phi} +
\gamma \wt{\phi}^2 R - \frac{1}{4} \wt{\phi}^2 f(-\ln \wt{\phi} )
F_{\mu\nu}^2 \right. \nn \\
&+& \left. \wt{\phi}^2  b(-\ln \wt{\phi} ) g^{\mu\nu}
\partial_{\mu} \psi \partial_{\nu} \psi + \wt{\phi}^2 V(-\ln
\wt{\phi}, \psi) \right].
\eea
Making the transformation
\beq
c_1\varphi = \gamma \wt{\phi}^2, \ \ g_{\mu\nu} = e^{\txs 2\rho} \,
\wt{g}_{\mu\nu}, \ \ \wt{\phi} = \left( \frac{c_1\varphi}{\gamma}
\right)^{1/2}, \ \ \rho = \frac{\gamma  \wt{\phi}^2}{4c_1^2} -
\frac{1}{8\gamma} \ln \wt{\phi},
\eeq
we can rewrite expression (3) in the form
\bea
S &=& \int d^2x \, \sqrt{\wt{g}} \left[ \frac{1}{2}
\wt{g}^{\mu\nu} \partial_{\mu} \varphi \partial_{\nu} \varphi +
c_1 \varphi \wt{R}
- \frac{1}{4} \exp\left(-\frac{\varphi }{2c_1} \right)
\left( \frac{c_1\varphi }{\gamma} \right)^{1+1/(8\gamma)}
f\left(- \frac{1}{2} \ln \frac{c_1\varphi }{\gamma} \right)
\wt{F}_{\mu\nu}^2 \right.  \nn \\
&+& \left. \frac{c_1\varphi }{\gamma} \,   b \left(- \frac{1}{2}
\ln
\frac{c_1\varphi }{\gamma} \right)  \wt{g}^{\mu\nu}
\partial_{\mu} \psi \partial_{\nu} \psi
+ \exp\left(\frac{\varphi }{2c_1} \right) \left(
\frac{c_1\varphi }{\gamma} \right)^{1-1/(8\gamma)} V(-
\frac{1}{2} \ln \frac{c_1\varphi }{\gamma} , \psi) \right],
\eea
with $\wt{F}_{\mu\nu}^2 = \wt{g}^{\mu\alpha} \wt{g}^{\nu\beta}
F_{\mu\nu} F_{\alpha\beta}$. Dropping the tilde for
simplicity and denoting again the Maxwell-dilaton and
scalar-dilaton couplings, and the dilaton potential in terms of
$\varphi$ as $f(\varphi )$, $b(\varphi )$ and $V(\varphi, \psi)$,
respectively, we get
\beq
S = \int d^2x \, \sqrt{g}  \left[ \frac{1}{2} g^{\mu\nu}
\partial_{\mu} \varphi \partial_{\nu} \varphi + c_1 \varphi
 R - \frac{1}{4} f(\varphi ) F_{\mu\nu}^2
+ b(\varphi ) g^{\mu\nu} \partial_{\mu} \psi
\partial_{\nu} \psi + V(\varphi, \psi) \right].
\eeq

Simple power counting shows that this theory is renormalizable in
the generalized sense (i.e., assuming a possible change of $f$, $b$
and
$V$ under renormalization). The perturbative renormalization of
 action (6) without the Maxwell term has been investigated in
refs. [12-15]. It has been shown there that the theory is
multiplicatively renormalizable for some choices of the potential
$V$. (In particular, it is well known that in the conformal gauge
dilaton gravity is renormalizable for $V=\Lambda $ or $V= \mu
e^{\alpha \varphi}$). Notice that, generally speaking, there
exist some gauges where dilaton gravity is not one-loop
renormalizable, or is a one-loop-finite theory [15]. The
perturbative properties and, in particular, the two-loop
finiteness of the supersymmetric extension of matter-dilaton
gravity  (i.e., two-dimensional supergravity) have recently been
investigated in ref. [21].

Our main purpose in this section will be to calculate the
contribution of the Maxwell sector to the one-loop counterterms
of the theory with action (6). Since the scalar field $\psi$ does
not interact with the Maxwell term, the background scalar $\psi$
can be put equal to zero in such calculation. Thus, the spectator
sector decouples from the Maxwell sector, and can be considered
independently. Hence, in this section we put $\psi =0$ and drop
all the $\psi$-dependence from the action (6). We are left with
\beq
S= \int d^2x \, \sqrt{g}  \left[ \frac{1}{2} g^{\mu\nu}
\partial_{\mu} \varphi \partial_{\nu} \varphi + c_1 \varphi
 R - \frac{1}{4} f(\varphi ) F_{\mu\nu}^2 + V(\varphi,
\psi) \right].
\eeq

The background field method will be used. According to this
procedure,
we split the fields into their quantum and background parts
\beq
g_{\mu\nu} \longrightarrow \bar{g}_{\mu\nu} =g_{\mu\nu}
+h_{\mu\nu}, \ \ \ \varphi \longrightarrow \bar{\varphi} = \phi +
\varphi, \ \ \ A_{\mu}  \longrightarrow \bar{A}_{\mu} =B_{\mu} +
A_{\mu},
\eeq
where the second terms $(h_{\mu\nu},\varphi,A_{\mu})$ are the
quantum fields.

Let us choose the gauge fixing actions in the gravitational and
electromagnetic field sectors, respectively, as the following
(linear covariant gauge)
\bea
S^g_{GF} &=& -\frac{c_1}{2} \int d^2x \, \sqrt{g} \left(
\nabla_{\nu}h^{\nu}_{\mu}- \frac{1}{2} \nabla_{\mu} h -
\frac{1}{\phi} \nabla_{\mu} \varphi \right) \phi \left(
\nabla_{\rho}h^{\rho\mu}- \frac{1}{2} \nabla^{\mu} h -
\frac{1}{\phi} \nabla^{\mu} \varphi \right), \label{gf1} \\
S^{A_{\mu}}_{GF} &=& -\frac{1}{2} \int d^2x \, \sqrt{g} f(\phi )
\left( \nabla_{\mu} A^{\mu} \right)^2. \label{gf2}
\eea
The gauge-fixing actions (\ref{gf1}) and
(\ref{gf2}) are to be added
to the quadratic expansion ($S^{(2)}$) of (7) on the quantum
fields.

Now let us recall a  few simple expressions that are necessary
for the one-loop counterterms calculation. The one-loop
divergences of the euclidean effective action are given by
\beq
\Gamma_{div} = -\frac{1}{2}  \tr \ln \left. \wh{\cal
H}\right|_{div}+  \tr \ln \left. \wh{\cal M}_{gh}\right|_{div},
\label{gdi}
\eeq
where $\wh{\cal H}$ is defined through $S^{(2)} + S_{GF}$, and
$\wh{\cal M}_{gh}$ is the ghost operator corresponding to the
gauge fixing action.

If  $\wh{\cal H}$ has the following form
\beq
\wh{\cal H} = \wh{1} \Delta + 2 \wh{E}^{\lambda} \nabla_{\lambda}
+ \wh{\Pi}, \label{hpe}
\eeq
where  $\wh{\cal H}$ acts in the space of all quantum fields (as
well as $ \wh{1}$, $ \wh{E}^{\lambda}$ and $ \wh{\Pi}$), and
$\Delta= \nabla^{\mu}\nabla_{\mu}$, then, in dimensional
regularization,
\beq
  \tr \ln \left. \wh{\cal H}\right|_{div} = \frac{1}{\varepsilon}
\tr \left( \wh{\Pi}- \wh{E}^{\lambda}\wh{E}_{\lambda}\right).
\eeq
Here, the parameter of dimensional regularization is $\varepsilon
=
2\pi (n-2)$. Notice that  we have dropped in (13) the
surface terms like $R$.
After the preceding remarks, we can now start the explicit
calculation
of the one-loop counterterms.
In order to simplify the calculus we will do it in two steps.
In the first one we will be interested in the contribution of
the Maxwell sector to the action of dilaton gravity (the first
two terms and $V$ of (7)). In this case we can put the
background vector field  $B_{\mu}=0$. Moreover, we can see
immediately that, due to the presence of only quadratic
terms in $A_{\mu}$, in $S^{(2)}$ the Maxwell sector decouples
from the gravitational sector. The corresponding term in the path
integral has the form (with account to the gauge fixing term
(10))
\beq
\int {\cal D} A^{\mu} \exp \left. \left( -\frac{1}{2} \int d^2x \,
\sqrt{g} f(\phi ) A_{\mu}
\wh{H}^{\mu\nu} A_{\nu} \right) \right|_{div} = \left. -
\frac{1}{2} \tr \ln \wh{H}^{\mu\nu} \right|_{div},
\eeq
where
\beq
 \wh{H}^{\mu\nu} = g^{\mu\nu} \Delta -  R^{\mu\nu}-
\frac{1}{f(\phi)} \left[ \delta_{\alpha}^{\mu} \left(
\nabla^{\nu} f(\phi) \right)- \delta_{\alpha}^{\nu} \left(
\nabla^{\mu} f(\phi) \right)- g^{\mu\nu} \left( \nabla_{\alpha}
f(\phi) \right) \right] \nabla^{\alpha},
\eeq
the local factor $f(\phi)$ not giving any contribution to
divergences. We see that the operator  $ \wh{H}^{\mu\nu}$
has exactly the form (\ref{hpe}), with
\beq
\wh{\Pi} = -R^{\mu\nu}, \ \ \ \wh{E}_{\alpha} = -
\frac{1}{2f(\phi)} \left[  \delta_{\alpha}^{\mu} \left(
\nabla^{\nu} f(\phi) \right)- \delta_{\alpha}^{\nu} \left(
\nabla^{\mu} f(\phi) \right)- g^{\mu\nu} \left( \nabla_{\alpha}
f(\phi) \right) \right].
\eeq
A simple calculation of the one-loop counterterms with this $
\wh{H}^{\mu\nu}$ shows that there is no non-trivial contribution
from the Maxwell sector to the dilaton gravity sector (except for
the
trivial surface terms which we consequently drop). The ghost
operator corresponding to the gauge fixing (10) is $\Delta$,
which again gives contribution  to the surface terms only. In
summary, the whole Maxwell sector does not provide any contribution
to
the quantum gravitational sector.

We proceed now with the calculation of the one-loop
counterterms corresponding to the Maxwell action ($B_{\mu} \neq
0$). For the sake
of simplicity, we can put $\phi=$ const. and  $g_{\mu\nu}=
\delta_{\mu\nu}$ in
the background field splitting (7). Note that the
calculation of the one-loop counterterms in the pure gravitational
sector, for the gauge under discussion, has been done already
[12,14] (see also the Appendix). It is not influenced by the
addition of
$F_{\mu\nu}$, as
we showed before; this is why we can choose $g_{\mu\nu}=
\delta_{\mu\nu}$.

First of all, we should write the quadratic expansion of the
action (7) with account to the corresponding gauge fixing
terms. A straightforward (although lengthy) calculation gives (for
the background under consideration)
\beq
S^{(2)} + S^g_{GF} +S^{A_{\mu}}_{GF}= \frac{1}{2} \int d^2x \,
\sqrt{g}
\ \left( A_{\rho} \ \
\bar{h}_{\mu\nu}  \ \ h \ \ \varphi \right) \wh{\cal H} \left(
\brr{c}
A_{\tau} \\ \bar{h}_{\alpha\beta} \\ h \\ \varphi \err \right),
\eeq
the 16 components of the matrix $\wh{\cal H}$ being given by
\bea
H_{11} &=& f(\phi ) g^{\rho\tau} \Delta,
H_{12} = -f(\phi ) \left[ B^{\alpha\rho} \nabla^{\beta}-
\delta^{\rho\beta} B^{\alpha\sigma} \nabla_{\sigma} \right]-
\frac{1}{2} f(\phi ) \left[ (\nabla^{\beta}B^{\alpha\rho}) -
\delta^{\rho\beta} ( \nabla_{\sigma}B^{\alpha\sigma})  \right], \nn
\\
H_{13} &=& -\frac{1}{2} f(\phi )  B^{\sigma\rho} \nabla_{\sigma}-
\frac{1}{4} f(\phi )  (\nabla_{\sigma}B^{\sigma\rho}), \ \ \ \
H_{14} =  f^{(1)}(\phi )  B^{\sigma\rho} \nabla_{\sigma}+
\frac{1}{2}
f^{(1)}(\phi )  (\nabla_{\sigma}B^{\sigma\rho}), \nn \\
H_{21} &=& f(\phi ) \left[ B^{\nu\tau} \nabla^{\mu}-
\delta^{\tau\mu} B^{\nu\sigma} \nabla_{\sigma} \right]- \frac{1}{2}
f(\phi ) \left[ (\nabla^{\mu}B^{\nu\tau}) -\delta^{\tau\mu} (
\nabla_{\sigma}B^{\nu\sigma})  \right], \nn \\
H_{22} &=&  \frac{1}{2} \left( c_1 \phi \Delta - \wt{V} \right)
\wh{P}^{\mu\nu, \alpha\beta}- f(\phi) B^{\mu\sigma}B^{\rho}_{\
\sigma}
\delta^{\nu\tau} \wh{P}^{\alpha\beta}_{\rho\tau}, \ \ \ H_{23}
=-\frac{f(\phi)}{4} B^{\mu\sigma} B^{\nu}_{\ \sigma} = H_{32},
 \nn \\
H_{24} &=&  \frac{f^{(1)}(\phi)}{2} B^{\mu \sigma} B^{\nu}_{\
\sigma }=H_{42}, \    \
H_{31} = \frac{1}{2} f(\phi )  B^{\sigma\tau} \nabla_{\sigma}-
\frac{1}{4} f(\phi )  (\nabla_{\sigma}B^{\sigma\tau}), \ \
H_{33}=0,
\nn \\
 H_{34} &=&  \frac{1}{2} \left(-c_1 \Delta + \wt{V}^{(1)}
+\frac{f^{(1)}(\phi)}{2} B_{\mu\nu}^2 \right) =H_{43},
  \\
H_{41} &=& -f^{(1)}(\phi )  B^{\sigma\tau} \nabla_{\sigma}+
\frac{1}{2}
f^{(1)}(\phi )  (\nabla_{\sigma}B^{\sigma\tau}),  \ \
H_{44} =   \left(\frac{c_1}{\phi}-1\right) \Delta +
2\wt{V}^{(2)}, \nn
\eea
where $\wh{P} \equiv \delta^{\mu\nu,\alpha\beta} -\frac{1}{2}
g^{\mu\nu} g^{\alpha\beta}$, $h_{\mu\nu}\equiv \bar{h}_{\mu\nu}
+\frac{1}{2} \delta_{\mu\nu} h$, $B_{\mu\nu} \equiv \nabla_{\mu}
B_{\nu}-\nabla_{\nu} B_{\mu}$, $\wt{V} \equiv V(\phi ) -\frac{1}{4}
f(\phi ) B_{\mu\nu}^2$, and $ \wt{V}^{(1)}$ and $ \wt{V}^{(2)}$ are
the first and second derivative of the potential with respect to
$\phi$, respectively, $\nabla_{\nu}$ being a flat derivative. The
contraction with the projector $\wh{P}$ of terms with $\alpha\beta$
and
$\mu\nu$ indices should also be done (because
$\bar{h}_{\alpha\beta} = \wh{P}_{\alpha\beta}^{\rho\tau}
\bar{h}_{\rho\tau}$,
$\bar{h}_{\mu\nu} = \wh{P}_{\mu\nu}^{\rho\tau}
\bar{h}_{\rho\tau}$).

One easily sees that the operator $\wh{\cal H}$ (18) does not have
the canonical structure (12) and, therefore, we cannot use the
algorithm (13). In order to make it possible the application
of this algorithm, we use a trick that was introduced in refs.
[12,14], namely we factorize
express
\beq
\wh{H} = \wh{K} \wh{\cal H}, \label{red}
\eeq
where $\wh{K}$ is the constant matrix:
\beq
\wh{K} = \left( \brr{cccc} \displaystyle
\frac{\delta^{\mu\nu}}{f(\phi
)} & 0 & 0 & 0 \\ 0 & \dsp \frac{2}{c_1\phi} \wh{P} & 0 & 0 \\
 0 & 0 & \dsp \frac{4}{c_1} \left( \dsp \frac{1}{c_1}- \dsp
\frac{1}{\phi}
\right) & \dsp - \frac{2}{c_1} \\ \displaystyle 0 & 0 & \dsp
-\frac{2}{c_1} & 0 \err \right). \eeq
The operator $\wh{H} $ ---which can be easily evaluated from
(\ref{red})--- has now actually the canonical structure (12). Owing
to the fact that  $\wh{K}$ is a constant matrix,
it is now immediate that we  have
\beq
\left. \tr \ln \wh{H} \right|_{div} = \left. \tr \ln \wh{\cal H}
\right|_{div},
\eeq
so that we can now concentrate ourselves on the operator $\wh{H}$
only.

Using the explicit form of  $\wh{H}$, we get
\beq
\wh{H} = \wh{1} \Delta + 2 \wh{E}^{\lambda} \nabla_{\lambda}
+ \wh{\Pi}, \label{hpe2}
\eeq
where the non-zero components of the matrix $E_{\lambda}$ are just
\bea
(E_{\lambda})_{12} &=&  -\frac{1}{2} \left[ B^{\alpha \rho}
\delta^{\beta}_{\lambda}- B^{\alpha}_{ \ \lambda}
\delta^{\rho\beta}
-\delta^{\alpha\beta} B_{\lambda}^{\ \rho}
\right], \ \ \ \
(E_{\lambda})_{13} =  -\frac{1}{4} B_{\lambda}^{ \ \rho}, \nn \\
(E_{\lambda})_{14} &=&   \frac{f^{(1)}(\phi )}{2 f(\phi )}
B_{\lambda}^{ \ \rho}, \  \ \ \
(E_{\lambda})_{21} =   \frac{f(\phi )}{c_1\phi } \left(
B^{\nu\tau} \delta^{\mu}_{\lambda}- B^{\nu}_{ \ \lambda}
\delta^{\tau\mu} -g^{\mu\nu} B^{ \ \tau}_{\lambda}  \right),
\label{ela}
\\ (E_{\lambda})_{31} &=&   \frac{1}{c_1} \left[
\left(\frac{1}{c_1}-
\frac{1}{\phi} \right) f(\phi )+f^{(1)} (\phi ) \right]
B_{\lambda}^{ \ \tau}, \ \ \ \
(E_{\lambda})_{41} =  -\frac{f(\phi )}{2 c_1} B_{\lambda}^{ \
\tau}. \nn
\eea
Concerning the operator $\wh{\Pi}$, we need only its diagonal
components for the calculation of divergences. They are
\bea
\Pi_{11} =0, & & \Pi_{22} =
-\frac{\wt{V}}{c_1\phi} \wh{P}^{\alpha\beta,\mu\nu}
-\frac{2f(\phi)}{c_1\phi} \wh{P}^{\alpha\beta}_{\mu'\nu'}
B^{\mu'\sigma} B^{\rho}_{\ \sigma} \delta^{\nu'\tau}
\wh{P}^{\mu\nu}_{\rho\tau}, \nn \\
& & \Pi_{33} =\Pi_{44} =-\frac{\wt{V}^{(1)}}{c_1}
-\frac{f^{(1)}(\phi)}{2c_1} B^2_{\mu\nu}.
\eea
It is now straightforward to apply the algorithm (10) and
to calculate the one-loop divergences:
\beq
\Gamma_{div} = -\frac{1}{2} \tr \ln \wh{H} \label{div1}
= \frac{1}{\varepsilon} \int d^2x \, \sqrt{g} \left\{
\frac{V(\phi )}{c_1\phi} +\frac{V^{(1)}(\phi )}{c_1} -\frac{1}{2}
B_{\mu\nu}^2 \left[ \frac{f(\phi )}{2c_1^2} +\frac{f^{(1)} (\phi
)}{2c_1} \right]\right\}.
\eeq
Notice, as is easy to see, that the ghosts corresponding to the
gauge fixing actions (9) and (10) give  contributions to
the surface terms only. Here, the expression (\ref{div1}) is the
final one for the one-loop divergences (Maxwell sector
contribution) of the effective action. We
can also observe that the one-loop counterterms for the potential
$V$
(which actually come from the dilaton-gravity sector)
coincide with the corresponding expressions found independently in
refs. [12,14]  (using the same gauge condition).

In order to write the final result we must also take into account
the counterterms of the gravitational sector (arbitrary
dilaton-graviton background), which interacts here with the Maxwell
sector. These counterterms have been
calculated in refs. [12,14]  (with the gauge fixing action (9), in
particular).
As has been shown above, the Maxwell sector does not
give here any contribution. In summary, all the one-loop
divergences
of the effective action for the theory (9) are\footnote{Due to the
absence of Maxwell sector contributions to the gravitational
counterterms, putting $B_{\mu\nu}=0$ we get the one-loop
divergences of dilaton gravity [12-14] (see Appendix).} (see
Appendix)
\beq
\Gamma_{div} =  \frac{1}{\varepsilon} \int d^2x \, \sqrt{g}
\left[ \frac{3}{\phi^2} \left(\nabla_{\lambda} \phi \right)
\left(\nabla^{\lambda} \phi \right)
+\frac{V(\phi )}{c_1 \phi} + \frac{V^{(1)}(\phi )}{c_1} -
\frac{1}{4} B_{\mu\nu}^2 \left( \frac{f(\phi )}{c_1^2}
+\frac{f^{(1)}(\phi )}{c_1} \right)\right]. \label{div2}
\eeq
This expression constitutes the main result of this section (see
also ref. [22]). It is interesting to note too, that the preceding
calculation is the two-dimensional analog of the
calculation done in ref. [23] for $d=4$ Maxwell-Einstein gravity.
\bs

\section{One-loop divergences: The scalar sector contributions.}

Let us now calculate the one-loop divergences which come from the
scalar sector interacting with gravity. Here, one can put
$F_{\mu\nu}=0$ in the action (6).

Once more, we split the fields into their quantum and background
parts:
\beq
g_{\mu\nu} \longrightarrow \bar{g}_{\mu\nu} =g_{\mu\nu}
+h_{\mu\nu}, \ \ \ \varphi \longrightarrow \bar{\varphi} = \phi +
\varphi, \ \ \ \psi  \longrightarrow \bar{\psi} =\psi +\eta,
\eeq
where the second terms $(h_{\mu\nu},\varphi,\eta)$ are the
quantum fields. In the gravitational sector we use again the
gauge-fixing action (9).

As in the previous section, to simplify the calculation we do it
 in two steps. In the first one we will be interested in the
scalar $\psi$ sector contribution to the action of dilaton
gravity. Once more, we can put here the background scalar $\psi
=0$.
Owing to the presence of quadratic terms in $\eta$ only, in
$S^{(2)}$
the scalar sector decouples from the gravitational sector. The
corresponding contribution to the effective action divergences is
\beq
 \int d\eta  \,  \exp \left[ -\int d^2x \, \sqrt{g}
b(\phi ) \eta \wh{H} \eta \right] = -\frac{1}{2}  \tr \ln \left.
\wh{H}\right|_{div},
\eeq
where $\wh{H} = \Delta + g^{\mu\nu} [\nabla_{\mu} b(\phi
)/b(\phi)] \partial_{\nu} - \ddot{V}(\phi, \psi)/b(\phi)$ and where
the dot denotes derivative with respect to $\psi$. A very simple
calculation, using (13), gives
\beq
 -\frac{1}{2}  \tr \ln \left. \wh{H}\right|_{div} =
\frac{1}{\epsilon} \int d^2x \, \sqrt{g} \left[ \frac{1}{8} \left(
\frac{b^{(1)}(\phi)}{b(\phi)} \right)^2 g^{\mu\nu} \partial_{\mu}
\phi  \partial_{\nu} \phi + \frac{\ddot{V} (\phi, \psi)}{2b(\phi)}
\right].
\eeq

Now we proceed with the calculation of the scalar-dilaton gravity
contribution to the terms with $\psi$ in the action (6). As in the
previous section, we can choose $g_{\mu\nu} = \delta_{\mu\nu}$,
$\phi =$ const., and arbitrary $\psi$ background for this
calculation. The quadratic expansion of the action (6) on this
background gives:
\beq
S^{(2)} + S^g_{GF} = \frac{1}{2} \int d^2x \, \sqrt{g}
\ \left( \eta \ \
\bar{h}_{\mu\nu}  \ \ h \ \ \varphi \right) \wh{\cal H} \left(
\brr{c}
\eta \\ \bar{h}_{\alpha\beta} \\ h \\ \varphi \err \right),
\eeq
being the components of $\wh{\cal H}$
\bea
H_{11} &=& -2b(\phi ) \Delta +2\ddot{V}, \ \
H_{12} =  b(\phi) (\nabla^{\alpha} \nabla^{\beta} \psi) +2 b(\phi)
(\nabla^{\alpha} \psi) \nabla^{\beta}, \nn \\
H_{13} &=& \frac{1}{2} \dot{V} =H_{31},  \ \ \ \
H_{14} = \dot{V}^{(1)} -b^{(1)} (\phi ) (\Delta \psi )-2
b^{(1)}(\phi )  ( \nabla^{\sigma} \psi)\nabla_{\sigma},  \nn \\
H_{21} &=&  b(\phi) (\nabla^{\nu} \nabla^{\mu} \psi) -2 b(\phi)
(\nabla^{\mu} \psi) \nabla^{\nu}, \nn \\
H_{22} &=& \frac{1}{2} \left( c_1 \phi \Delta - \wt{V} \right)
\wh{P}^{\mu\nu, \alpha\beta}+ 2b(\phi) (\nabla^{\mu} \psi)
(\nabla^{\alpha} \psi) \delta^{\nu\beta},
 \\ H_{23}
&=& \frac{b(\phi)}{2} (\nabla^{\mu} \psi) (\nabla^{\nu} \psi)
=H_{32},
  \ \
H_{24} = -b^{(1)}(\phi) (\nabla^{\mu} \psi) (\nabla^{\nu} \psi)
=H_{42}, \nn \\
H_{33} &=& 0, \ \ \
 H_{34} =  \frac{1}{2} \left( -c_1 \Delta + \wt{V}^{(1)}
-b^{(1)}(\phi) (\nabla^{\mu} \psi) (\nabla_{\mu} \psi) \right)
=H_{43},     \nn \\
H_{41} &=&  \dot{V}^{(1)} -b^{(1)} (\phi ) (\Delta \psi )+2
b^{(1)}(\phi )  ( \nabla^{\sigma} \psi)\nabla_{\sigma}, \ \ \
H_{44} =   \left(\frac{c_1}{\phi}-1\right) \Delta +
2\wt{V}^{(2)}, \nn
\eea
where $\wt{V} = V +b(\phi) (\nabla^{\mu} \psi) (\nabla_{\mu}
\psi)$ and multiplication with the projector $\wh{P}$ of the terms
with indices $\alpha\beta$ and $\mu\nu$ is assumed.
Splitting $ \wh{\cal H}$ in the form (19), the only component
of $\wh{K}$ which changes in (20) is $K_{11}$:
\[
K_{11}= -\frac{1}{2b(\phi )},
\]
and we get the operator $\wh{H}$ in the form (22) with the
following $\wh{E}_{\lambda}$ and diagonal components of $\wh{\Pi}$
\bea
& & \Pi_{11} =-\frac{\ddot{V}}{b(\phi )}, \ \ \ \Pi_{22} =
-\frac{\wt{V}}{c_1\phi} \wh{P}^{\mu\nu,\alpha\beta}
+\frac{4b(\phi)}{c_1\phi} \wh{P}^{\mu\nu}_{\mu'\nu'}
(\nabla^{\mu'}\psi )(\nabla^{\alpha'}\psi ) \delta^{\nu'\beta'}
\wh{P}^{\alpha\beta}_{\alpha'\beta'}, \nn \\
& & \Pi_{33} =\Pi_{44} =-\frac{\wt{V}^{(1)}}{c_1}
+\frac{b^{(1)}(\phi)}{c_1} g^{\mu\nu} \partial_{\mu} \psi
\partial_{\nu} \psi;
\eea
the non-zero components of $\wh{E}_{\lambda}$ being
\bea
 & & (E_{\lambda})_{12} =-\frac{1}{2}
\wh{P}^{\alpha\beta}_{\sigma\lambda} (\nabla^{\sigma} \psi ), \ \
\ \  (E_{\lambda})_{14} =\frac{b^{(1)}(\phi )}{2b(\phi )}
(\nabla_{\lambda} \psi ), \nn \\
& & (E_{\lambda})_{21} =-\frac{2b(\phi )}{c_1\phi }
\wh{P}^{\mu\nu}_{\sigma\lambda} (\nabla^{\sigma} \psi ), \ \ \
 (E_{\lambda})_{31} =-\frac{2b^{(1)}(\phi )}{c_1} (\nabla_{\lambda} \psi
).
\eea
Using the algorithm (13) with (32), (33), one can now find the
scalar-gravitational contribution to the effective action
divergences on the constant dilaton background:
\beq
\Gamma_{div} = \frac{1}{\varepsilon} \int d^2x \, \sqrt{g} \left[
\frac{V(\phi, \psi )}{c_1\phi} +\frac{V^{(1)}(\phi, \psi )}{c_1}+
\frac{\ddot{V} (\phi, \psi )}{2b(\phi )} \right].
\eeq
The first two terms in expression (34) come from the
dilaton-gravity sector (and have been already taken into account in
eq. (26)). The third term has been also evaluated in (29) (when
$\psi =0$).  Hence, we have a partial checking of our calculation
above. It is interesting to notice that there is some miraculous
cancellation of the possible counterterm for $b(\phi )(\nabla
\psi)^2$.
Finally, we  conclude that the only new divergent terms which
appear in the scalar sector are given by expression (29). Thus, the
final result for the one-loop divergences of the effective action
of the theory (6) is the following (the sum of (26) and (29)):
\bea
\Gamma_{div} &=& \frac{1}{\varepsilon} \int d^2x \, \sqrt{g}
\left\{
\left[ \frac{3}{\phi^2} + \frac{1}{8} \left( \frac{b^{(1)}(\phi
)}{b(\phi )} \right)^2 \right] (\nabla_{\lambda} \psi )
(\nabla^{\lambda} \psi ) +
\frac{V(\phi, \psi )}{c_1\phi} \right. \nn \\
 &+& \left. \frac{V^{(1)}(\phi, \psi )}{c_1}+ \frac{\ddot{V} (\phi,
\psi )}{2b(\phi )} -\frac{1}{4} B_{\mu\nu}^2 \left( \frac{f(\phi
)}{c_1}+ \frac{f^{(1)} (\phi )}{c_1} \right)  \right\}.
\eea
This expression constitutes the main result of our work.
\bs

\section{One-loop renormalization.}

Let us now discuss the
renormalization of the theory under consideration. Taking into
account the one-loop divergences just calculated (expression (35)),
the one-loop renormalized action is
\bea
S_R &=& \int d^2x \, \sqrt{g} \left\{ \frac{1}{2} \left[ 1-
\frac{6}{\varepsilon \varphi^2}- \frac{1}{4\varepsilon} \left(
\frac{b^{(1)}(\varphi )}{b(\varphi )}
\right)^2  \right] g^{\mu\nu} \partial_{\mu}
\varphi \partial_{\nu} \varphi +c_1 \varphi R+ V(\varphi, \psi )
\left(1- \frac{1}{\varepsilon c_1 \varphi} \right) \right. \nn \\
& -& \left. \frac{V^{(1)} (\varphi, \psi )}{\varepsilon c_1}
 - \frac{\ddot{V} (\varphi, \psi )}{2\varepsilon c_1}
-\frac{f(\varphi )}{4} F_{\mu\nu}^2 \left( 1- \frac{1}{\varepsilon
c_1^2} -\frac{f^{(1)}(\varphi )}{\varepsilon c_1 f(\varphi )}
\right) +b(\varphi )  g^{\mu\nu} \partial_{\mu}
\psi \partial_{\nu} \psi \right\}.
\eea
Choosing the one-loop renormalization of fields and the constant
$c_1$
in the dilaton gravity sector as the following (see [12,14] for a
discussion of this point)
\beq
\varphi   =\varphi_R, \ \ \ \ c_{1} = c_{1R}, \ \ \ \ g_{\mu\nu}
= e^{\txs 2\sigma (\varphi, c_1, \varepsilon)} \, \wt{g}_{\mu\nu},
\eeq
where, in the one-loop approximation, $\sigma$ satisfies the
following equation
\beq
\partial_{\nu} \sigma (\varphi, c_1, \varepsilon )
=\frac{1}{4c_1\varepsilon} \left[ 6\partial_{\nu} \varphi^{-1}-
\frac{1}{4} \left(\frac{b^{(1)}(\varphi )}{b(\varphi )}
\right)^2 \partial_{\nu} \varphi \right],
\eeq
 we obtain:
\bea
S_R &=& \int d^2x \, \sqrt{\wt{g}} \left[ \frac{1}{2}
\wt{g}^{\mu\nu}
\partial_{\mu} \varphi \partial_{\nu} \varphi +c_1 \varphi \wt{R}+
V(\varphi, \psi ) \left(1- \frac{1}{\varepsilon c_1 \varphi} +
2\sigma (\varphi, c_1, \varepsilon) \right)
- \frac{V^{(1)} (\varphi, \psi )}{\varepsilon c_1}
\right. \nn \\
& -& \left.
\frac{\ddot{V} (\varphi, \psi )}{2\varepsilon c_1}
-\frac{f(\varphi )}{4} \wt{F}_{\mu\nu}^2 \left( 1-
\frac{1}{\varepsilon c_1^2} -2\sigma (\varphi, c_1, \varepsilon)
 -\frac{f^{(1)} (\varphi
)}{\varepsilon c_1 f(\varphi )} \right)+ b(\varphi )  \wt{g}^{\mu\nu}
\partial_{\mu}
\psi \partial_{\nu} \psi \right].
\eea
Note that the scalar $\psi $ does not get renormalized in the
one-loop approximation (as is the case with $\varphi$).

One can see that the theory is one-loop renormalizable in the
generalized sense (i.e., admitting possible changes of $f$ and $V$
under renormalization). Moreover, as we will now show, the theory
(6) can be one-loop multiplicatively renormalizable
for some choices of the functions $V$, $b$ and $f$.

First, let us choose
\beq
V(\varphi, \psi )= V_1(\varphi) e^{\sqrt{a(\varphi )} \, \psi}+
V_2(\varphi) e^{-\sqrt{a(\varphi )} \, \psi}.
\eeq
Then, the term involving $\ddot{V}$ becomes
\beq
V(\varphi, \psi ) \left(-\frac{a (\varphi )}{2\varepsilon c_1}
\right).
\eeq
(Notice that if $V(\varphi, \psi )=V(\varphi ) (1+\psi )$, then the
term in  $\ddot{V}$ is equal to zero in expression (39)).
We can actually exhibit some examples of functions $V$, $a$, $b$
and $f$ for which the theory is multiplicatively renormalizable (in
the gauges (9), (10)), namely the following.
\ben
\item $b(\varphi )=b=$ const.,  $a(\varphi )=$ const. Then
\bea
f(\varphi ) = \frac{f_0}{\varphi^3} \exp \left[ -\left(
\frac{1}{c_1}+ \wt{b} \right) \varphi \right], & &
f_0= \left( 1- \frac{ \wt{b}}{\varepsilon c_1} \right) f_0^R, \nn
\\
V(\varphi ) = \mu \varphi^2 \exp \left[ -\left( \frac{a}{2}+ \wt{a}
\right) \varphi \right], & &
\mu = \left( 1- \frac{ \wt{a}}{\varepsilon c_1} \right) \mu_R,
\eea
where $\wt{a}$ and $\wt{b}$ are arbitrary constants. Notice that
for  $\wt{a}=\wt{b}=0$ the theory is one-loop finite.
\item $b(\varphi )=e^{4b\varphi} $,  $a(\varphi )=a$,\, with $a,b$
const.
Then
\bea
f(\varphi ) = \frac{f_0}{\varphi^3} \exp \left[ -\left(
\frac{1}{c_1}+ \wt{b}-2b^2 \right) \varphi \right], & &
f_0= \left( 1- \frac{ \wt{b}}{\varepsilon c_1} \right) f_0^R, \nn
\\
V(\varphi ) = \mu \varphi^2 \exp \left[ -\left( \frac{a}{2}+
\wt{a}+2b^2 \right) \varphi \right], & &
\mu = \left( 1- \frac{ \wt{a}}{\varepsilon c_1} \right) \mu_R,
\eea
Observe that here the renormalization of $f_0$ and $\mu$ is the
same as in (42). This is also true for the following case.
\item  $b(\varphi )=b_0+b_1\varphi $,  $a(\varphi )=a$,\, with
$a,b_0,
b_1$ const. We have
\bea
& & f(\varphi ) = \frac{f_0}{\varphi^3}(b_0+b_1\varphi)^{-1/8} \exp
\left[ -\left( \frac{1}{c_1}+ \wt{b} \right) \varphi \right],
 \nn \\
& & V(\varphi ) = \mu \varphi^2 (b_0+b_1\varphi)^{1/8} \exp \left[
-\left( \frac{a}{2}+ \wt{a} \right) \varphi \right].
\eea
It can be seen that the choice of $b(\varphi )$ and  $a(\varphi )$
is a crucial one. For different choices of these functions,
different choices of $V$ and $f$ (from (39)) lead to a one-loop
renormalizable theory. We will use the expressions (42)-(44) for
the discussion of the corresponding charged black holes in the next
section.
\een

Let us now compare the one-loop structure of the theory (6)
with the structure of four-dimensional Maxwell-Einstein theory,
which is known to be one-loop non-re\-nor\-mal\-iz\-able [23], even
on
shell. First of all, one can see that there is the contribution to
the gravitational sector ($R^2$-terms) from the electromagnetic
sector in four dimensions [23]. As we have seen, in the
two-dimensional case there is no contribution to the
dilaton-gravity sector. Second, while in four spacetime dimensions
there is no contribution from
the gravitational sector to $F_{\mu\nu}^2$, we see
here, on the contrary, that for Maxwell-dilaton gravity
there is such contribution from the dilaton-gravity sector, even if
the function $f(\varphi )$ is chosen to be constant. Third, in
ordinary spacetime new
counterterms appear [23] which differ from the original action and
which, in the end, lead to the non-renormalizability of
four-dimensional Maxwell-Einstein theory even on shell. Such terms
have not appeared in the theory (6), which is one-loop
multiplicatively renormalizable off-shell for $V$ and $f$ given by
(42) or (43).
This suggests that the theory under discussion here is more similar
to quantum
$R^2$-gravity with matter (see the book [26] for a general review),
which is known to be multiplicatively renormalizable and even
asymptotically free for some gauge groups [26]. However, in
$R^2$-gravity with matter (as for the Einstein-Maxwell theory)
there is a contribution to the gravitational sector coming from the
electromagnetic sector and there is no gravitational sector
contribution to $F_{\mu\nu}^2$.

Finally, we must point out that for the
two-dimensional theory under consideration ---as well  as for the
four-dimensional Maxwell-Einstein theory--- one-loop counterterms
(off-shell) depend on the gauge. Here, presumably, the use of a
gauge independent effective action (see [24,25] for a discussion of
gauge independent effective action in two-dimensional gravity) can
add some information on the renormalization structure of the
theory.
\bs

\section{Equations of motion and charged black holes.}

The equations of motion for our general action (1) are easily
obtained:
\bea
2 \nabla^2 \phi -4 (\nabla \phi )^2+ \frac{1}{4} f(\phi )
F_{\mu\nu}^2+
 V(\phi,\psi)&=&0, \nn \\
R_{\mu\nu}+2 \nabla_{\mu}\nabla_{\nu} \phi +(\gamma -4) \left[
\nabla_{\mu} \phi \nabla_{\nu} \phi- g_{\mu\nu} (\nabla \phi
)^2+\frac{1}{2} g_{\mu\nu} \nabla^2 \phi \right]
 - \frac{1}{2} f(\phi ) F_{\mu}^{ \ \lambda} F_{\nu\lambda}
\ \ \label{emo} & &
\\
+ \frac{1}{16} f^{(1)}(\phi ) g_{\mu\nu} F_{\alpha\beta}^2+ b(\phi
)
\nabla_{\mu}
\psi\nabla_{\nu} \psi -\frac{1}{4} b^{(1)}(\phi ) g_{\mu\nu}
(\nabla
\psi)^2 -\frac{1}{4}  g_{\mu\nu} \frac{\partial
V(\phi,\psi)}{\partial \phi} &=&0, \nn \\
\nabla^{\mu} \left[ F_{\mu\nu} f(\phi ) e^{-2\phi} \right] \, = \,
0,
\ \ \
2b(\phi ) \nabla^2 \psi + 2 [b^{(1)}(\phi )-2b(\phi )] \nabla \phi
\cdot
\nabla \psi - \frac{\partial V(\phi,\psi)}{\partial \psi}&=&0, \nn
\eea
where $f^{(1)}$ means first derivative with respect to $\phi$.
In the case of static spherical configurations (also considered
in
[18]): \ $F=\wt{f}(r)dr\wedge dt$, $\phi (r)$, $\psi (r)$, \,
with
an asymptotically flat metric \ $ds^2 =-g(r)dt^2+ g(r)^{-1}
dr^2$,
and \, $g(r)\rightarrow 1$ as $r \rightarrow \infty$,  we obtain:
\bea
(g\phi')'-2g(\phi')^2- \frac{f}{4} \wt{f}^2+
\frac{1}{2} V (\phi,\psi) =0, \ \ \ \ \
2\phi'' + (\gamma -4) (\phi')^2+b(\psi')^2 &=& 0, \nn \\
\left( f \wt{f} e^{-2\phi} \right)' =0, \ \ \ \ \
(g\psi')'+ \left( \frac{b^{(1)}}{b}-2\right) g\phi' \psi'-
\frac{1}{2b}
\, \frac{\partial V (\phi,\psi)}{\partial \psi} &=&0, \label{emo1}
\eea
where the prime means derivative with respect to $r$.

We shall restrict ourselves to the first of the solutions obtained
before (the one corresponding to eq. (43)). By expanding in
$e^{-\phi}$,
the potential $V$
is seen to correspond to a special case of the type of dilaton
potential
one expects from closed string loop corrections [17]. In fact, in
terms of the original dilaton field $\phi$ (see (4)), we get
\beq
V(\phi )=  \mu' \, e^{\txs \alpha' \phi} \, e^{\txs A'e^{\txs
-2\phi}}, \ \ \
\ \ \ f(\phi )=  f' \, e^{\txs -(\alpha'-6) \phi}\ e^{ \txs
B'e^{\txs -2\phi}}, \ \ \ \ \ \  b(\phi)= b\, e^{2\phi}, \label{pot}
\eeq
or, for some choice of $\wt{b}, \wt{a}$,
\beq
V(\phi )=  \mu' \, e^{\txs \alpha' \phi},
 \ \ \
\ \ \ f(\phi )=  f' \, e^{\txs -(\alpha'-6) \phi}.
\eeq

These are indeed very nice functions of the variable $\phi$. In
particular, for positive values of all the
 coefficients involved, $V(\phi )$ in (47) develops
minimum
for a rather small value of $\phi$ (depending, of course, on the
precise
values of the constants), while both $f$ and  $b$  have
monotonic
exponential shapes (see Fig. 1). When $A'$ and $B'$ are negative,
$V$ is
an exponentially increasing function of $\phi$ while $f$ has a
maximum
previous to an exponential decrease (Fig. 2). $V$ also develops a
minimum for
negative values of $\mu'$, $\alpha$ and $A'$. Moreover, notice that
if
one uses the expansion on $\phi$ of $\exp (\exp (-2\phi))$ (as in
the
perturbative case of refs. [17,18]), then {\it the minimum is never
seen}.

We look now for the charged black holes of our theory.
\ms

\subsection{Particular case $\psi=$ const.}

To warm up, in a first instance our
emphasis ---in comparing with [17,18]--- will be put in the
fact of allowing the Maxwell-dilaton coupling
 to be an arbitrary function of
the field $\phi$ and
---for the sake of simplicity of the discussion--- we shall not
take into account (for the moment) the dependence on the scalar
spectator field $\psi$.

 The solutions of the
two last eqs. (\ref{emo}) (when $\psi$ is put equal to constant)
are [18]
\beq
\wt{f}(r)=\wt{f}_0 f^{-1} (\phi (r)) e^{2\phi (r)}, \ \ \ \
\phi (r )= \left\{
\brr{ll}
\phi_0 + \frac{\alpha}{2} \ln r,  &  \mbox{for} \ \gamma\neq 4,
\\ \phi_0 - \frac{Q}{2} \, r,   & \mbox{for} \ \gamma = 4, \err
\right. \eeq
where $\alpha \equiv 4/(\gamma -4)$, and
$\wt{f}_0$, $\phi_0$ and $Q$ are arbitrary constants (see also
[18]).
The
solution for $g(r)$ (coming from the first of eqs. (4)) is
\beq
g(r)= \left\{
\brr{ll}
r^{\alpha +1} \left[ -2m- \frac{1}{\alpha} \int^r ds \, s^{-\alpha}
W(\phi (s)) \right], & \mbox{for} \ \gamma\neq 4, \\
e^{-Qr} \left[ -2m+ \frac{1}{Q} \int^r ds \, e^{Qs}
W(\phi (s)) \right], & \mbox{for} \ \gamma = 4, \err \right.
\label{gr1}
\eeq
where $W(\phi )\equiv V(\phi ) -\wt{f}_0^2e^{4\phi}/(2f(\phi))$ and
$m$
is a new constant, which turns out to be the black hole's mass.

This solution for the potential
(\ref{pot})
(below, we change slightly
(\ref{pot}) by choosing the signs of the parameters in order to
correspond to what we expect from closed strings loops,
$e^{B'e^{2\phi}}$,
$e^{A'e^{2\phi}}$,
 and
to obtain the asymptotic condition $g(r) \rightarrow 1$, $r\rightarrow
\infty$)
 describes a generalized
Reissner-Nordstr{\o}m black hole. In this case, Figs. 1 and 2 are
replaced by the (very similar ones) Figs. 3 and 4, respectively. An
explicit expression
for the metric $g(r)$ can be obtained by direct substitution of the
particular form of the potential chosen
 into the preceding expression (\ref{gr1}). Let us consider first the
case $\gamma \neq 4$ and suppose that $\alpha <0$ (this last is no
restriction for all meaningful situations). The two integrals which
appear in the calculation of $g(r)$ have the same general form.
Take the first one
\beq
I_1 \equiv \int_0^r ds \, s^{\alpha (-1+\alpha'/2)} \, \exp (-
ks^{\alpha}).
\eeq
The change of variables: $ks^{\alpha} =u$, yields
\beq
I_1 =-\frac{1}{\alpha} k^{1-1/\alpha-\alpha'/2}
\int_{kr^{\alpha}}^{\infty} du \, u^{1/\alpha+\alpha'/2-2} \, e^u
=-\frac{1}{\alpha} k^{1-1/\alpha-\alpha'/2} \, \Gamma_{kr^{\alpha}}
\left( \frac{1}{\alpha} +\frac{\alpha'}{2 }-1 \right).
\eeq
An analogous result is obtained for the second integral (which is
actually the dominant one for the asymptotic behaviour  $r
\rightarrow \infty$). It is now easy to see that the asymptotic
condition $g(r)\rightarrow 1$ as $r \rightarrow \infty$ can be
fulfilled provided that
\beq
\alpha' = 2-\frac{4}{\alpha}.
\eeq
So, in particular, for $\gamma =2,3,-4$ ---which correspond to
 $\alpha =-2,-4,-1/2$--- we have
$\alpha' =4,3,10$, respectively.

Let us now turn to the special case $\gamma =4$. The integrals are
then of the form
\beq
I_1' =\int_0^r ds \, e^{(1-\alpha'/2)Qs} \, e^{ke^{-Qs}},
\eeq
and have the following asymptotic behaviour for  $r \rightarrow
\infty$:
\beq
I_1'(r) \sim  e^{(1-\alpha'/2)Qr}, \ \ \ \ I_2'(r) \sim  e^{(2-
\alpha'/2)Qr},
\eeq
what yields the asymptotic condition $g(r)\rightarrow 1$ as $r
\rightarrow \infty$, provided that $\alpha'=2$ (again, the second
term is the dominant one for this asymptotic behaviour).

It is important to remark that our dilaton potential has the
general form to be expected from closed string loop corrections
(power series in $e^{2\phi}$). The explicit form of the metric
$g(r)$ can be given in terms of incomplete gamma functions and has
a consistent asymptotic behaviour for large $r$. Our solution
constitutes a new explicit example (actually a whole family of
them) of the spacetime geometries associated with multiple horizons
(see [17]) and the fact that it is of the same form as the one which
satisfies
the renormalizability requeriments makes it particularly interesting.
\ms

\subsection{General case with arbitrary spectator field $\psi$.}

Let us now turn to the more difficult situation in which we take into
account
the dependences on the field $\psi$. The whole set of eqs.
(\ref{emo1}) has to be solved. The fourth one is simple, it just
says that
\beq
 \psi (r)=c_1 \int
\frac{dr}{g(r)} + c_2,
\eeq
and the third one reduces in this general case to
\beq
\wt{f}(r)= \frac{c_3}{f'} \, e^{\txs (\alpha'-4) \phi (r)}\, e^{\txs -B'
e^{\txs 2\phi (r)}}.
 \eeq
Also $g(r)$ can be put in terms of the field $\phi (r)$, by using
the second eq. (\ref{emo1})
\beq
g(r)= \frac{c_1\sqrt{b} e^{\phi (r)} }{\left[ (4-\gamma)
(\phi')^2(r)-2 \phi'' (r)\right]^{1/2}}.
\eeq
Finally, we are left with a single diferential equation for the
field $\phi (r)$. It reads
\bea
&&\frac{\phi'''(r)\phi'(r)-2 (\phi'')^2 (r) }{\left[ (4-\gamma)
(\phi')^2(r)-2 \phi'' (r)\right]^{3/2}}-
\frac{ (\phi')^2 (r) }{\left[ (4-\gamma) (\phi')^2(r)-2 \phi''
(r)\right]^{1/2}} \nn \\
&-& \frac{c_3^2}{4c_1\sqrt{b} f'} \,  e^{(\alpha' -3)\phi (r)} \,
e^{-B'e^{2\phi (r)}} + \frac{\mu'}{2c_1\sqrt{b}} \,  e^{(\alpha' -
1)\phi (r)} \,  e^{A'e^{2\phi (r)}} =0. \label{gfe}
\eea
Although involving one single function of one variable, $\phi (r)$,
this equation is very difficult to solve. We can try some
convenient limits, as for instance the case when $b \rightarrow
\infty$, which leads to the much simpler equation:
\beq
\phi'''(r)\phi'(r)-2 (\phi'')^2 (r)+2 (\phi')^2 (r) \phi'' (r) +
(\gamma -4) (\phi')^4 (r) =0.
\eeq
This equation has the trivial solution $\phi =c=$ const. and, for
$\gamma =4$, also $\phi = ar+b$, $a,b=$ const. In general, the
asymptotic behaviour for $r \rightarrow \infty$ is easily seen to
be the following:
\beq
\phi (r) \sim -2 \ln \left( \frac{r}{r_0} \right),
\eeq
which leads, in particular, to $g(r) \rightarrow 1$,  $r
\rightarrow \infty$, thus providing a further check of our procedure.
Solving (\ref{gfe}) and substituting into the three preceding
equations we get the general solution of our model.
A more detailed analysis of the solutions to these equations involves
rather technical mathematical reasoning and will be given elsewhere.
 \bs

\section{Conclusions.}

In the present paper we have discussed the issue of one-loop
renormalization in two-dimensional dilaton quantum gravity with matter
described by a scalar and abelian gauge field. We have shown that the
theory is not only renormalizable in the generalized sense, but also
that it
is actually one-loop multiplicatively renormalizable for some choices of
the dilaton potential and of the dilaton-Maxwell and dilaton-scalar
couplings.  In fact, the model under discussion is the two-dimensional
analog of the quantum gravity-Maxwell-scalar system in four dimensions
(which is nonrenormalizable for Einstein gravity, or  is nonunitary
for higher-derivative gravity). Hence, the properties of the
two-dimensional model are much better ---the theory is renormalizable
and unitary. We have also studied some properties of the classical
field equations, in particular, the static solutions which describe
charged black holes.

It is interesting to remark that, actually, we do not have direct
analogs of the usual couplings in dilaton-matter gravity (except for
$c_1$, whose corresponding $\beta$-function is trivially zero).
Concerning this point, it would be important to discuss now
dilaton-matter
gravity (for example, gravity-Yang-Mills theory), for which there are
some non-trivial coupling constants, as the gauge coupling constant.
Then, it would be also rewarding to find the effective coupling
constant in order to check if such important properties as, for
example, asymptotic freedom can be met in 2D matter-dilaton gravity.

\vspace{5mm}

\ni{\large \bf Acknowledgments}

We would like to thank Profs. Ignatios Antoniadis, Ron Kantowski
and
Andrei Slavnov for fruitful discussions on connected problems, and
Ali Chamseddine and Arkadii Tseytlin for correspondence.
S.D.O. thanks the members of the Department E.C.M. of
Barcelona University for their warm hospitality.
This work has
been supported by Direcci"n Ge\-ne\-ral de Investigaci"n
Cient!fica y Tcnica (DGICYT), research projects
 PB90-0022 and SAB92-0072.
\bs

\appendix

\section{Appendix}

In this appendix we describe the calculation of the one-loop
counterterms for pure dilaton gravity (action (6) with
$A_{\mu}=\psi
=0$). We follow ref. [14].
First of all, it is necessary to expand the action in terms of the
quantum fields
$(h_{\mu\nu}, \varphi)$ on the arbitrary background
$(g_{\mu\nu}, \phi)$. We substitute in this expansion the
two-dimensional
identity
\beq
R_{\alpha\beta} =\frac{1}{2} g_{\alpha\beta} R.
\eeq
As is explained in ref. [14], use of this identity does not give
the contribution to the pole $1/(n-2)$ (except for the surface
terms, which we consequently drop). A straightforward (though
extremely tedious) calculation gives (in the gauge (9))
\beq
S^{(2)} + S^g_{GF} = \frac{1}{2} \int d^2x \, \sqrt{g}
\ \left( \bar{h}_{\mu\nu}  \ \ h \ \ \varphi \right) \wh{\cal H}
\left(
\brr{c}
 \bar{h}_{\alpha\beta} \\ h \\ \varphi \err \right),
\eeq
where $\wh{H} =\wh{K} \wh{\cal H}$, $\wh{K} $ is a $3\times 3$
matrix (namely, $\wh{K} $ is given by the matrix (20) without the
first row and first column), and the operator $\wh{H} $ has the
form (22). The non-zero components of the operator
$\wh{E}_{\lambda}$ are
\bea
(E^{\lambda})_{11} &=& \frac{1}{\phi} \left[ g^{\nu\beta}
g^{\lambda\alpha} (\nabla^{\mu} \phi ) -  g^{\nu\beta}
g^{\lambda\mu} (\nabla^{\alpha} \phi ) -\frac{3}{2}
\delta^{\mu\nu,\alpha \beta} (\nabla^{\lambda} \phi ) \right], \nn
\\
(E^{\lambda})_{12} &=& -\frac{1}{\phi}(\nabla^{\mu} \phi )
g^{\nu\lambda} =(E^{\lambda})_{21}, \ \ \ (E^{\lambda})_{13} = -
\frac{1}{c_1\phi}(\nabla^{\mu} \phi ) g^{\nu\lambda},  \\
(E^{\lambda})_{23} &=& \frac{1}{\phi^2}(\nabla^{\lambda} \phi ),
\ \ \ \  (E^{\lambda})_{31} = -\frac{1}{2}(\nabla^{\beta} \phi )
g^{\alpha\lambda}. \nn
\eea
The diagonal components of the operator $\wh{\Pi}$ are
\bea
\Pi_{11} &=&  -\frac{4}{\phi}
g^{\nu\beta}(\nabla^{\mu}\nabla^{\alpha} \phi )+ \frac{4}{c_1\phi}
g^{\mu\alpha} \left[ \frac{1}{2}(\nabla^{\nu} \phi )
(\nabla^{\beta} \phi )+ c_1\phi R^{\nu\beta} \right]  \\
&-& \frac{1}{c_1\phi} \delta^{\mu\nu,\alpha\beta} \left[
\frac{1}{2}(\nabla \phi )^2+ c_1\phi R + V \right], \ \ \
\Pi_{22} =  - \frac{\phi -c_1}{c_1\phi} \, (\Delta \phi), \ \ \
\Pi_{33}  =   - \frac{2}{c_1}  \, V^{(1)}. \nn
\eea
Using the algorithm (13), we obtain
\beq
 -\frac{1}{2}  \tr \ln \left. \wh{H}\right|_{div} =
\frac{1}{\varepsilon} \int d^2x \, \sqrt{g} \left[
\frac{V}{c_1\phi}
+ \frac{V^{(1)}}{c_1}
 + \frac{17}{4\phi^2}(\nabla \phi)^2 \right]. \label{65}
\eeq
The ghost operator corresponding to the gauge-fixing action (9) is
\beq
-\wh{\cal M}_{gh} = \delta_{\alpha}^{\beta} \Delta - \frac{1}{\phi}
(\nabla^{\beta} \phi ) \nabla_{\alpha}+ R_{\alpha}^{\ \beta}  -
\frac{1}{\phi} (\nabla_{\alpha} \nabla^{\beta} \phi ). \label{66}
\eeq
{}From (13) and (\ref{66}), we get
\beq
 \tr \ln \wh{\cal M}_{gh} = \frac{1}{\varepsilon} \int d^2x \,
\sqrt{g} \left[ - \frac{5}{4\phi^2} (\nabla \phi)^2 \right]. \label{67}
\eeq
Finally, by addition of (\ref{65}) and (\ref{67}) we obtain the
divergences of the effective action for pure dilaton gravity
\beq
\Gamma_{div} = \frac{1}{\varepsilon} \int d^2x \, \sqrt{g}
\left[ \frac{3}{\phi^2} g^{\alpha\beta} (\partial_{\alpha} \phi )
(\partial_{\beta} \phi ) + \frac{1}{c_1\phi} V + \frac{1}{c_1}
V^{(1)} \right]. \label{68}
\eeq
It follows from the above calculations that the two last terms in
(\ref{68}) (which have been obtained in sections 2 and 3 also) originate
from the dilaton gravity sector.

\newpage




\newpage

\ni{\large \bf Figure captions}.
\bs

\ni{\bf Fig. 1}. For positive values of the constants, the potential
$V(\phi) $ of (\ref{pot}) develops a nice minimum for a particular value
 $\phi_0$ (which depends, of course, on the specific values of the
parameters). This minimum is {\it never seen} in the perturbative
expansion of $V(\phi)$ (refs. [17,18]).
 $V(\phi)$  also develops a minimum for negative
values
of $\mu'$, $\alpha$ and $A'$.
\ms

\ni{\bf Fig. 2}. For $A'$ and $B'$ negative, the function $f(\phi)$ of
(\ref{pot}) shows a maximum, previous to an exponential decrease
(it takes over the role $V(\phi)$
had for positive values of these parameters).
 \ms

\ni{\bf Fig. 3}. Same as Fig. 1 but for the potential (\ref{pot})
modified (sign of uppermost exponent), now being  $\alpha' <0$ and
$A'>0$. \ms

\ni{\bf Fig. 4}. As Fig. 3,   for  $\alpha' >0$ and $A'<0$.

\end{document}